\documentclass[aps,prb,twocolumn,groupedaddress,showpacs]{revtex4}
\usepackage{graphicx}
\usepackage{wasysym}
\usepackage{bm}

\begin{document}
\title{Stacking-order dependence in thermoelectric transport of biased trilayer graphene}
\author{R. Ma$^1$}
\email{njrma@hotmail.com}
\author{L. Sheng$^2$}
\email{shengli@nju.edu.cn}
\author{M. Liu$^3$}
\author{D. N. Sheng$^4$}
\address{$^1$ School of Physics and Optoelectronic Engineering,
Nanjing University of Information Science and Technology, Nanjing
210044, China\\
$^2$ National Laboratory of Solid State Microstructures and
Department of Physics, Nanjing University, Nanjing 210093, China\\
$^3$ Department of Physics, Southeast University, Nanjing 210096,
China\\
$^4$ Department of Physics and Astronomy, California State
University, Northridge, California 91330, USA}

%\date{\today}

\begin{abstract}
We numerically study the thermoelectric and thermal transport in
trilayer graphene with different stacking orders in the presence of
interlayer bias under a strong perpendicular magnetic field. In
biased ABA-stacked case, we find that the thermoelectric
conductivity displays different asymptotic behaviors with the
varying of the temperature, similar to that of monolayer graphene.
In the high temperature regime, the transverse thermoelectric
conductivity $\alpha_{xy}$ saturates to a universal value $2.77 k_B
e/h$ at the center of each LL, while it displays a linear
temperature dependence at low temperatures limit. The calculated
transverse thermal conductivity $\kappa_{xy}$ exhibits two plateaus
away from the band center. The transition between the two plateaus
is continuous, which is accompanied by a pronounced peak in the
longitudinal thermal conductivity $\kappa_{xx}$. In biased
ABC-stacked case, it is found that both the thermoelectric
conductivity and thermal conductivity have similar properties to the
biased bilayer graphene, which is consistent with the behavior of a
band insulator. The obtained results demonstrate the sensitivity of
the thermoelectric conductivity to the band gap near the Dirac
point. We also verify the validity of the Mott-relation and the
generalized Wiedemann-Franz law.

\end{abstract}

\maketitle

\section{Introduction}
\label{sec:intro}

Recently, much attention has been paid to the thermoelectric
transport properties of graphene both experimentally~\cite{Kim09,Shi09,Ong08} and
theoretically~\cite{CastroNeto09,DasSarma09,Fogelstrom07,Thalmeier07,Ting09}.
In experiments, the thermopower (the longitudinal
thermoelectric response) and the Nernst signal (the transverse
response) in the presence of a strong magnetic field are found to be
large, reaching the order of the quantum limit $k_B/e$, where $k_B$
and $e$ are the Boltzmann constant and the electron charge,
respectively~\cite{Kim09,Shi09,Ong08}. Besides monolayer graphene,
bilayer graphene is also very interesting. Experiments have
shown that bilayer graphene exhibits tunable
bandgap~\cite{zhang09,Kuzmenko09} in the presence of an applied
bias voltage, exhibiting similar properties to semiconductors.
Thermoelectric measurement~\cite{Hao10} shows that the
room-temperature thermopower with a bias voltage can be enhanced by
a factor of 4 compared to that of the monolayer graphene or
unbiased bilayer graphene, making it a more promising candidate for
future thermoelectric applications. Theoretical
calculations from the tight-binding
models for monolayer and bilayer graphene~\cite{zhu10,Ma11}
are in agreement with the experimental
observations~\cite{Kim09,Shi09,Ong08,Hao10,Lee10}.

More recently, the focus of the study of graphene systems has
gradually extended to trilayer
graphene~\cite{Taychatanapat11,zhang2011,Kumar11,Jhang11,Sena11,Henriksen2012,Koshino2010,Yuan2011}.
In trilayer graphene, the interlayer coupling in stacked layers of
graphene gives rise to even richer electronic transport properties.
Experimental and theoretical studies show
that~\cite{Guinea06,Min08,Koshino09,Avetisyan10}, the electronic
structure and the Landau level (LL) spectrum at the vicinity of the
Dirac point are very sensitive to the stacking order of the graphene
layers. Trilayer graphene has two stable stacking orders: (1) ABA
(Bernal) stacking, where the atoms of the topmost layer lie exactly
on top of those of the bottom layer; and (2) ABC (rhombohedral)
stacking, where atoms of one sublattice of the top layer lie above
the center of the hexagons in the bottom layer. This seemingly small
distinction in stacking order results in a dramatic difference in
band structures. The low-energy band structure for ABA-stacked
trilayer graphene contains both linear and hyperbolic bands, similar
to the combined spectrum of  monolayer graphene and  bilayer
graphene~\cite{Guinea07,Koshino07}, while ABC-stacked case presents
approximately cubic bands~\cite{Koshino09}. Moreover, the LL
spectrum of ABA-stacked case in a perpendicular magnetic field $B$
can be viewed as a superposition of $\sqrt{B}$-dependent
monolayer-like LLs and $B$-dependent bilayer-like
LLs~\cite{Guinea06,McCann06,Ezawa07}. On the other hand, the LLs of
ABC-stacked case are given by $E_n\propto B^{3/2}\sqrt{n(n-1)(n-2)}$
with Berry's phase 3$\pi$~\cite{Guinea06,Min08,Koshino09}.
Interestingly, when a bias voltage or a potential difference is
applied to the top and bottom graphene layers, ABA-stacked case
exhibits a semi-metallic band structure with a tunable band overlap
between the conduction and valence
bands~\cite{Craciun09,Koshino2009}, whereas ABC-stacked case
exhibits a semiconducting band structure with a tunable band gap,
similarly to bilayer
graphene~\cite{Guinea06,Koshino09,Avetisyan10,Bao11}. Owing to their
distinctive band structures, ABA- and ABC-stacked trilayer graphene
are expected to exhibit rich novel thermoelectric transport
properties. However, theoretical understanding of the thermoelectric
transport properties of trilayer graphene is limited compared to
that of monolayer or bilayer graphene. In particular, the influence
of different stacking orders on the thermoelectric transport
properties has not been studied so far, which is highly desired.

In this paper, we carry out a numerical study of the thermoelectric
transport properties in both ABA- and ABC-stacked trilayer graphene
systems in the presence of electrostatic bias between the top and
bottom graphene layers. The effects of disorder and thermal
activation on the broadening of LLs are considered. In biased
ABA-stacked case, the thermoelectric coefficients exhibit unique
characteristics near the central LL due to the LL crossing of
electron and hole bands, which are quite different from those of
biased bilayer graphene. Both the longitudinal and the transverse
thermoelectric conductivities are universal functions of the
effective bandwidth and temperature,
%% (e.g. $W_L/E_F$ and $k_BT/E_F$, see Fig.\ref{fig.2} for details),
and display different asymptotic behaviors in different temperature
regimes. The Nernst signal displays a peak at the central LL with a
height of the order of $k_B/e$, and changes sign near other LLs,
while the thermopower behaves in an opposite manner. The peak values
of the Nernst signal and thermopower are very large, compared with
monolayer graphene due to the semi-metallic band overlap near zero
energy. The validity of the semiclassical Mott relation is found to
remain valid at low temperatures. In biased ABC-stacked case, we
observe quite different behavior from biased ABA-stacked case near
the central LL. Around the Dirac point, the transverse
thermoelectric conductivity exhibits a pronounced valley at low
temperatures. This is attributed to the opening of a sizable gap
between the valence and conduction bands in biased ABC-stacked case.
In addition, we have calculated the thermal transport coefficients
of electrons for both biased ABA- and ABC-stacked trilayer graphene
systems. In biased ABA-stacked case, the calculated transverse
thermal conductivity $\kappa_{xy}$ exhibits two plateaus away from
the band center. The transition between the two plateaus is
continuous, which is accompanied by a pronounced peak in the
longitudinal thermal conductivity $\kappa_{xx}$. In biased
ABC-stacked case, the transverse thermal conductivity $\kappa_{xy}$
displays an apparent plateau with $\kappa_{xy}=0$, which is
accompanied by a valley in $\kappa_{xx}$, which provides an
additional evidence for the band insulator behavior. We further
compare the calculated thermal conductivities with those deduced
from the Wiedemann-Franz law, to check the validity of this
fundamental relation in trilayer graphene systems.

This paper is organized as follows. In Sec.\ II, we introduce the
model Hamiltonian. In Sec.\ III and Sec.\ IV, numerical results
based on exact diagonalization and thermoelectric transport
calculations are presented for biased ABA-stacked and ABC-stacked
trilayer graphene systems, respectively. In Sec.\ V, numerical
results for thermal transport coefficients are presented. The final
section contains a summary.

\begin{figure}[tbp]
\par
\includegraphics[height=0.21\textwidth]{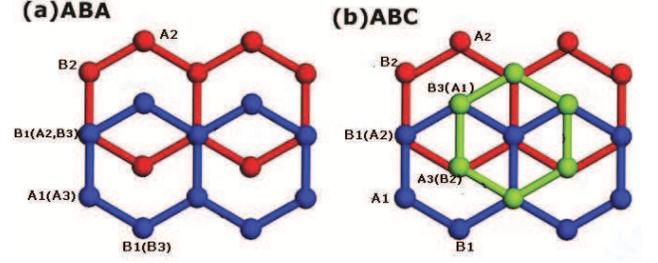}
\caption{(color online). Schematic of trilayer graphene lattice with
ABA and ABC stacking, where the blue/red/green lines indicate links
on the bottom/middle/top layers.} \label{fig.1}
\end{figure}

\section{Model and Methods}
\label{sec:model}

We consider a trilayer graphene system consisting of three coupled
hexagonal lattices including inequivalent sublattices $A_1$, $B_1$
on the bottom layer, $A_2$, $B_2$ on the middle layer, and $A_3$,
$B_3$ on the top layer. The three graphene layers are arranged in
the ABA (Bernal) or ABC (rhombohedral) stacking
orders~\cite{Dresselhaus}, as shown in Fig.\ref{fig.1}. The
difference between ABA and ABC stacking is the top layer. For ABA
stacking, the top layer will be exactly above the bottom layer
without any relative shift. For ABC stacking, $A_3$ sublattice of
the top layer lies above the centers of the hexagons in the bottom
layer, $B_3$ sublattice lies above the centers of the hexagons in
the middle layer. Here, the in-plane nearest-neighbor hopping
integral between $A_i$ and $B_i$ atoms is denoted by
$\gamma_{A_iB_i} =\gamma_{0}$ with $i=1,\cdots,3$. For the
interlayer coupling, we take into account two largest hopping
integrals. For ABA stacking, the largest interlayer hopping is
between a $B_1$ ($B_3$) atom and the nearest $A_2$ atom
$\gamma_{B_1A_2}=\gamma_{B_3A_2}=\gamma_{1}$. The smaller hopping is
between a $B_2$ atom and three nearest $A_1$ ($A_3$) atoms
$\gamma_{B_2A_1}=\gamma_{B_2A_3}=\gamma_{3}$. For ABC stacking, the
largest interlayer hopping is between a $B_1$ ($B_2$) atom and the
nearest $A_2$ ($A_3$) atom
$\gamma_{B_1A_2}=\gamma_{B_2A_3}=\gamma_{1}$. The smaller hopping is
between a $B_2$ ($B_3$) atom and three nearest $A_1$ ($A_2$) atoms
$\gamma_{B_2A_1}=\gamma_{B_3A_2}=\gamma_{3}$. The values of these
hopping integrals are taken to be $\gamma_{0}=3.16$ eV,
$\gamma_{1}=0.39$ eV, and $\gamma_{3}=0.315$ eV~\cite{Ma08}.

We assume that each monolayer graphene has totally $L_{y}$ zigzag
chains with $L_{x}$ atomic sites on each zigzag
chain~\cite{Sheng06}. The size of the sample will be denoted as
$N=L_{x}\times L_{y}\times L_{z}$, where $L_{z}=3$ is the number of
monolayer graphene planes along the $z$ direction. We have confirmed
that the calculated results does not depend on the system sizes (as
long as the system lengths are reasonably large)~\cite{Ma08}. In the
presence of an applied magnetic field perpendicular to the plane of
biased trilayer graphene, the lattice model for ABA stacking in real
space can be written in the tight-binding form:

\begin{eqnarray}
H&=&-\gamma_{0}(\sum\limits_{\langle
ij\rangle\sigma}e^{ia_{ij}}c_{1i\sigma}^{\dagger
}c_{1j\sigma}+\sum\limits_{\langle
ij\rangle\sigma}e^{ia_{ij}}c_{2i\sigma}^{\dagger}c_{2j\sigma}\nonumber\\
&+&\sum\limits_{\langle
ij\rangle\sigma}e^{ia_{ij}}c_{3i\sigma}^{\dagger }c_{3j\sigma})
-\gamma_{1}(\sum\limits_{\langle
ij\rangle_1\sigma}e^{ia_{ij}}c_{1j\sigma
B_1}^{\dagger}{c}_{2i\sigma{A_2}}\nonumber\\
&+&\sum\limits_{\langle ij\rangle_1\sigma}e^{ia_{ij}}c_{3j\sigma
B_3}^{\dagger}{c}_{2i\sigma{A_2}}) -\gamma_{3}(\sum\limits_{\langle
ij\rangle_3\sigma}e^{ia_{ij}}c_{2j\sigma
B_2}^{\dagger}{c}_{1i\sigma{A_1}} \nonumber\\
&+&\sum\limits_{\langle ij\rangle_3\sigma}e^{ia_{ij}}c_{2j\sigma
B_2}^{\dagger}{c}_{3i\sigma{A_3}})+h.c.
+\sum\limits_{i\sigma}w_{i}(c_{1i\sigma}^{\dagger }c_{1i\sigma}\nonumber\\
&+&{c}_{2i\sigma}^{\dagger }{c}_{2i\sigma}+{c}_{3i\sigma}^{\dagger
}{c}_{3i\sigma})+\sum\limits_{i\sigma}
(\epsilon_1c_{1i\sigma}^{\dagger}c_{1i\sigma}+\epsilon_2c_{3i\sigma}
^{\dagger}c_{3i\sigma}),
\end{eqnarray}
where $c_{mi\sigma}^{\dagger}$ ($c_{mi\sigma A_m}^{\dagger}$),
$c_{mj\sigma}^{\dagger}$ ($c_{mj\sigma B_m}^{\dagger}$) are creation
operators on $A_m$ and $B_m$ sublattices in the $m-th$ layer
($m=1,\cdots,3$), with $\sigma$ as a spin index. The sum
$\sum_{\langle ij\rangle\sigma}$ denotes the intralayer
nearest-neighbor hopping in three layers, $\sum_{\langle
ij\rangle_1\sigma}$ stands for the interlayer hopping between the
$B_1$ ($B_3$) sublattice in the bottom (top) layer and the $A_2$
sublattice in the middle layer, and $\sum_{\langle
ij\rangle_3\sigma}$ stands for the interlayer hopping between the
$B_2$ sublattice in the middle layer and the $A_1$ ($A_3$)
sublattice in the bottom (top) layer, as described above. For the
biased system, the top and the bottom graphene layers gain different
electrostatic potentials, and the corresponding energy difference is
given by $\Delta _g=\epsilon_2-\epsilon_1$ where
$\epsilon_1=-\frac{1}{2}\Delta_g$, and
$\epsilon_2=\frac{1}{2}\Delta_g$. For illustrative purpose, a
relatively large asymmetric gap $\Delta_g=0.3\gamma_0$ is assumed.
$w_{i}$ is a random disorder potential uniformly distributed in the
interval $w_{i}\in \lbrack -W/2,W/2]\gamma_0$. The magnetic flux per
hexagon $\phi =\sum_{{\small {\mbox{\hexagon}}}}a_{ij}=\frac{2\pi
}{M}$ is proportional to the strength of the applied magnetic field
$B$, where $M$ is assumed to be an integer and the lattice constant
is taken to be unity.

For ABC-stacked trilayer graphene in the presence of bias voltage,
the Hamiltonian can be written as:
\begin{eqnarray}
H&=&-\gamma_{0}(\sum\limits_{\langle
ij\rangle\sigma}e^{ia_{ij}}c_{1i\sigma}^{\dagger
}c_{1j\sigma}+\sum\limits_{\langle
ij\rangle\sigma}e^{ia_{ij}}c_{2i\sigma}^{\dagger}c_{2j\sigma}\nonumber\\
&+&\sum\limits_{\langle
ij\rangle\sigma}e^{ia_{ij}}c_{3i\sigma}^{\dagger }c_{3j\sigma})
-\gamma_{1}(\sum\limits_{\langle
ij\rangle_1\sigma}e^{ia_{ij}}c_{1j\sigma
B_1}^{\dagger}{c}_{2i\sigma{A_2}}\nonumber\\
&+&\sum\limits_{\langle ij\rangle_1\sigma}e^{ia_{ij}}c_{2j\sigma
B_2}^{\dagger}{c}_{3i\sigma{A_3}}) -\gamma_{3}(\sum\limits_{\langle
ij\rangle_3\sigma}e^{ia_{ij}}c_{2j\sigma
B_2}^{\dagger}{c}_{1i\sigma{A_1}} \nonumber\\
&+&\sum\limits_{\langle ij\rangle_3\sigma}e^{ia_{ij}}c_{3j\sigma
B_3}^{\dagger}{c}_{2i\sigma{A_2}})+h.c.
+\sum\limits_{i\sigma}w_{i}(c_{1i\sigma}^{\dagger }c_{1i\sigma}\nonumber\\
&+&{c}_{2i\sigma}^{\dagger }{c}_{2i\sigma}+{c}_{3i\sigma}^{\dagger
}{c}_{3i\sigma})+\sum\limits_{i\sigma}
(\epsilon_1c_{1i\sigma}^{\dagger}c_{1i\sigma}+\epsilon_2c_{3i\sigma}
^{\dagger}c_{3i\sigma}),
\end{eqnarray}
The sum $\sum_{\langle ij\rangle\sigma}$ denotes the intralayer
nearest-neighbor hopping in three layers, $\sum_{\langle
ij\rangle_1\sigma}$ stands for the interlayer hopping between the
$B_1$ ($B_2$) sublattice in the bottom (middle) layer and the $A_2$
($A_3$) sublattice in the middle (top) layer, and $\sum_{\langle
ij\rangle_3\sigma}$ stands for the interlayer hopping between the
$B_2$ ($B_3$) sublattice in the middle (top) layer and the $A_1$
($A_2$) sublattice in the bottom (middle) layer, as described above.

In the linear response regime, the charge current in response to an
electric field and a temperature gradient can be written as  ${\bf J}
= {\hat \sigma} {\bf E} + {\hat \alpha} (-\nabla T)$, where ${\hat
\sigma}$ and ${\hat \alpha}$ are the electrical and thermoelectric
conductivity tensors, respectively.  The transport coefficient
$\sigma_{xx}$ can be calculated by Kubo formula and $\sigma _{xx}$
can be  obtained based on the calculation of the Thouless number \cite{Ma08}.
In practice, we  first calculate the $T=0$ conductivities
$\sigma_{ji}(E_F)$, and then use the relation~\cite{Jonson84}
\begin{eqnarray}
\sigma_{ji}(E_F, T) &=& \int d\epsilon \,\sigma_{ji}(\epsilon)
\left ( - {\partial f(\epsilon) \over \partial \epsilon } \right), \nonumber \\
\alpha_{ji}(E_F, T) &=& {-1\over eT} \int d\epsilon\,
\sigma_{ji}(\epsilon) (\epsilon-E_F) \left ( - {\partial f(\epsilon)
\over \partial \epsilon } \right), \label{eq:conductance-finiteT}
\end{eqnarray}
to obtain the finite-temperature electrical and thermoelectric
conductivity tensors. Here, $f(x) = 1/[e^{(x-E_F)/k_B T}+1]$ is the
Fermi distribution function. At low temperatures, the second
equation can be approximated as
\begin{equation}
\alpha_{ji}(E_F, T) =-\frac {\pi^2k_B^2T}{3e}\left. \frac
{d\sigma_{ji}(\epsilon, T)}{d\epsilon} \right|_{\epsilon =E_F},
\label{eq:Mott-relation}
\end{equation}
which is the semiclassical Mott relation~\cite{Jonson84,Oji84}. The
thermopower and Nernst signal can be calculated subsequently
from~\cite{footnote1}
\begin{eqnarray}
S_{xx} &=&  { E_x \over \nabla_x T} =  {\rho_{xx}\alpha_{xx}
- \rho_{yx}\alpha_{yx}}, \nonumber \\
S_{xy} &=& { E_y \over \nabla_x T} =  {\rho_{xx}\alpha_{yx}
+\rho_{yx}\alpha_{xx}}.
\label{eq:thermoelectric}
\end{eqnarray}

The thermal conductivity, measuring the magnitude of the thermal
currents in response to an applied temperature gradient, which usually includes
electron and phonon contributions. In our numerical calculations,
phonon-related thermal conductivity is omitted. The electronic
thermal conductivities $\kappa_{ji}$ at finite temperature assume
the forms~\cite{Oji84}
\begin{eqnarray}
\kappa_{ji}(E_F, T) &=& {1\over e^2T} \int d\epsilon\,
\sigma_{ji}(\epsilon) (\epsilon-E_F)^2 \left ( - {\partial
f(\epsilon) \over \partial \epsilon} \right) \nonumber\\
&-&T\alpha_{ji}(E_F, T)\sigma_{ji}^{-1}(E_F, T)\alpha_{ji}(E_F, T).
\label{eq:thermal conductivity}
\end{eqnarray}
For diffusive electronic transport in metals, it is established that
the Wiedemann-Franz law is satisfied between the electrical
conductivity $\sigma$ and the thermal conductivity $\kappa$ of
electrons~\cite{Ziman}:
\begin{equation}
{\kappa\over \sigma T}=L , \label{eq:W-F law}
\end{equation}
where L is the Lorentz number and takes a constant value:
$L={\pi^2\over 3}({k_B\over e})^2$. The validity of this
relation will be examined for the present trilayer graphene.

\begin{figure*}[tbh]
\par
\includegraphics[width=0.8\textwidth]{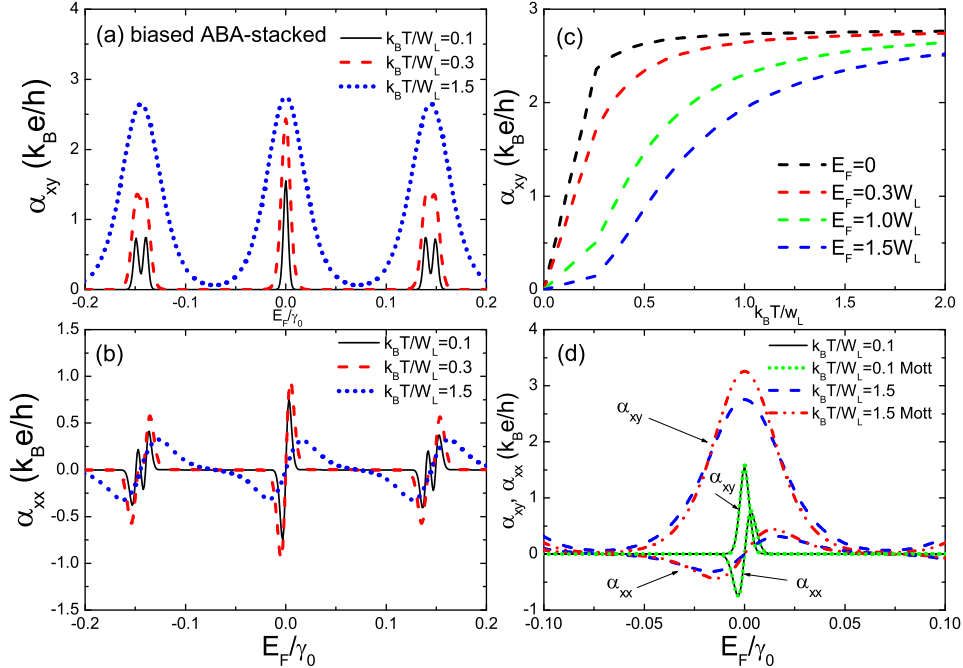}
\caption{ (color online). Thermoelectric conductivities at finite
temperatures of biased ABA-stacked trilayer graphene. (a)-(b)
$\alpha_{xy}(E_F, T)$ and $\alpha_{xx}(E_F,T)$ as functions of the
Fermi energy $E_F$ at different temperature $T$. (c) shows the
temperature dependence of $\alpha_{xy}(E_F,T)$ for trilayer
graphene. (d) Comparison of the results from numerical calculations
and from the generalized Mott relation at two characteristic
temperatures, $k_{B}T/W_L=0.1$ and $k_BT/W_L=1.5$. Here, the width
of the central LL $W_L/\gamma_0=0.0069$. The asymmetric gap
$\Delta_g=0.3\gamma_0$. The system size is taken to be
$N=96\times24\times3$, the magnetic flux $\phi=2\pi/48$, and the
disorder strength $w=0.1$. } \label{fig.2}
\end{figure*}

\begin{figure*}[tbh]
\par
\includegraphics[width=0.7\textwidth]{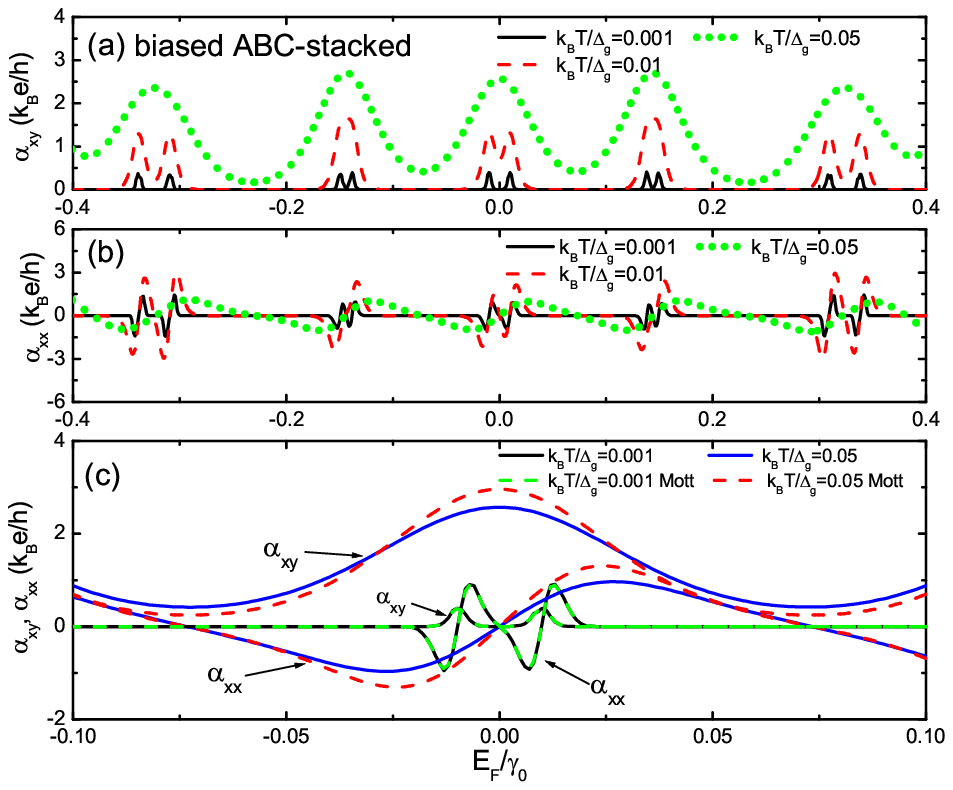}
\caption{ (color online). Thermoelectric conductivities at finite
temperatures of biased ABC-stacked trilayer graphene. (a)-(b)
$\alpha_{xy}(E_F, T)$ and $\alpha_{xx}(E_F,T)$ as functions of Fermi
energy at different temperatures. (c) Comparison of the results from
numerical calculations and from the generalized Mott relation at two
characteristic temperatures, $k_{B}T/\Delta_g=0.001$ and
$k_BT/\Delta_g=0.05$. The asymmetric gap $\Delta_g=0.3\gamma_0$. The
other parameters are chosen to be the same as in Fig. 2. }
\label{fig.3}
\end{figure*}

\section{Thermoelectric transport in biased ABA-stacked trilayer graphene systems}
\label{sec:ABA stacking}

We first show the calculated thermoelectric conductivities  at
finite temperatures for biased ABA-stacked trilayer graphene. As
shown in  Fig.\ref{fig.2}(a) and (b), the transverse thermoelectric
conductivity $\alpha_{xy}$ displays a series of peaks, while the
longitudinal thermoelectric conductivity ${\alpha_{xx}}$ oscillates
and changes sign at the center of each LL. At low temperatures, the
peak of $\alpha_{xy}$ at the central LL is higher and narrower than
others, which indicates that the impurity scattering has less effect
on the central LL. These results are qualitatively similar to those
found in monolayer graphene, but some  differences exist due to the
overlap of electron and hole bands. At low temperatures, more
oscillations are observed in the higher LLs than monolayer graphene,
in consistent with the further lifting of the LL degeneracy in
biased ABA-stacked case. As shown in Fig.\ref{fig.2}(b), around zero
energy, the peak value of $\alpha_{xx}$ shows different trends with
increasing temperature. It first increases with $T$ in the
low-temperature region, and then decreases with $T$ at high
temperatures. This is due to the competition between $\frac
{\pi^2k_B^2T}{3e}$ and $\frac {d\sigma_{ji}(\epsilon,
T)}{d\epsilon}$ of Eq.(\ref{eq:Mott-relation}). The peak value of
$\alpha_{xx}$ could either increase or decrease depending on the
relative magnitudes of these two terms. At high temperatures,
$\sigma_{ji}(\epsilon, T)$ becomes smooth, and consequently
$\alpha_{xx}$ begins to decrease. In Fig.\ref{fig.2}(c), we find
that $\alpha_{xy}$ shows different behavior depending on the
relative strength of the temperature $k_{B}T$ and the width of the
central LL $W_L$ ($W_L$ is determined by the full-width at the
half-maximum of the $\sigma_{xx}$ peak). When $k_{B}T \ll W_L$ and
$E_F \ll W_L$, $\alpha_{xy}$ shows linear temperature dependence,
indicating that there is a small energy range where extended states
dominate, and the transport falls into the semi-classical
Drude-Zener regime. When $E_F$ is shifted away from the Dirac point,
the low-energy electron excitation is gapped due to Anderson
localization. When $k_{B}T$ becomes comparable to or greater than
$W_L$, $\alpha_{xy}$ for all LLs saturates to a constant value $2.77
k_B e/h$. This matches exactly the universal value $(\ln 2) k_B e/h$
predicted for the conventional integer quantum Hall effect (IQHE)
systems in the case where thermal activation
dominates~\cite{Jonson84, Oji84}, with an additional degeneracy
factor $4$. The saturated value of $\alpha_{xy}$ in biased
ABA-stacked case is in accordance with the fourfold degeneracy at
zero energy. In the presence of bias voltage, the valley degeneracy
of the LLs usually is lifted by the interlayer potential asymmetry,
so that the 12-fold energy levels (four and eight levels from the
monolayer-like and the bilayer-like subbands, respectively) split
into six different levels with twofold spin degeneracy. However,
near the Dirac point, the interlayer potential asymmetry causes
hybridization of the linear and parabolic chiral bands, which leads
to the  fourfold degeneracy for zero energy Landau
levels.~\cite{Henriksen2012,Koshino2010,Yuan2011}

To examine the validity of the semiclassical Mott relation, we
compare the above results with those calculated from
Eq.(\ref{eq:Mott-relation}), as shown in Fig.\ref{fig.2}(d). The
Mott relation is a low-temperature approximation and predicts that
the thermoelectric conductivities have linear temperature
dependence. This is in agreement with our low-temperature results,
which proves that the semiclassical Mott relation is asymptotically
valid in the Landau-quantized systems, as suggested in
Ref.~\onlinecite{Jonson84}.

\section{Thermoelectric transport in biased ABC-stacked trilayer graphene systems}
\label{sec:ABC stacking}

For biased ABC-stacked trilayer graphene, we show the calculated
$\alpha_{xx}$ and $\alpha_{xy}$ at finite temperatures in
Fig.\ref{fig.3}. As seen from Fig.\ref{fig.3}(a), $\alpha_{xy}$
displays a pronounced valley at low temperature, in striking
contrast to ABA-stacked case with a peak at $E_F=0$. These results
are qualitatively similar to those found in biased bilayer
graphene~\cite{Ma11}. This behavior can be understood as due to the
split of the valley degeneracy in the central LL by an opposite
voltage bias added to the top layer and the bottom layer. This is in
consistent with the opening of a sizable gap between the valence and
conduction bands in biased ABC-stacked trilayer
graphene~\cite{Lui11}. ${\alpha_{xx}}$ oscillates and changes sign
around the center of each split LL. In Fig.\ref{fig.2}(c), we also
compare the above results with those calculated from the
semiclassical Mott relation using Eq.(\ref{eq:Mott-relation}). The
Mott relation is found to remain valid at low temperatures.

\begin{figure*}[tbp]
\par
\includegraphics[width=0.8\textwidth]{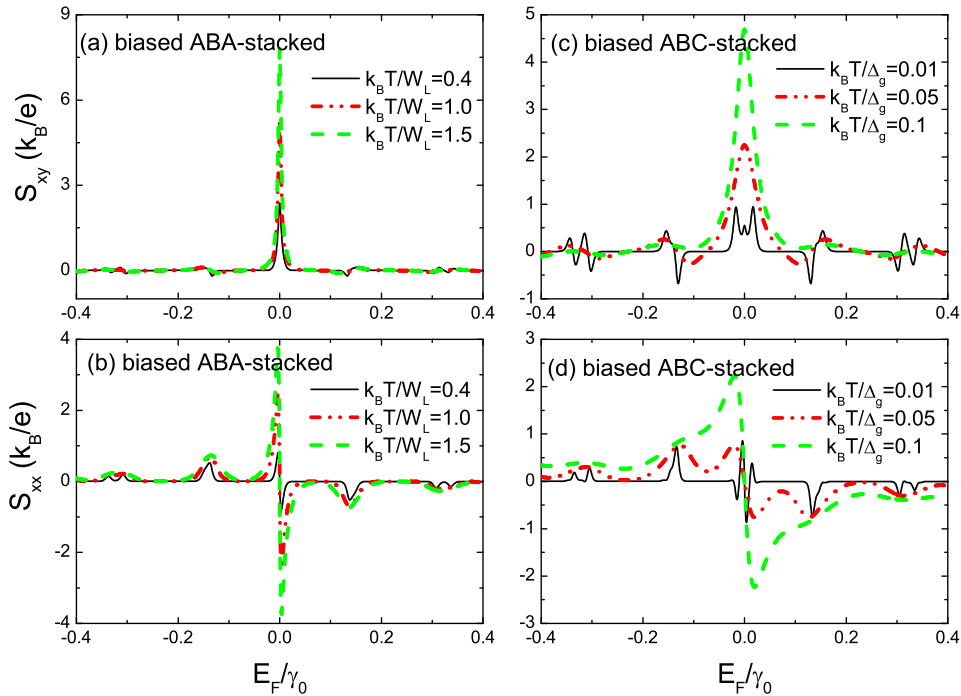}
\caption{(color online). The thermopower $S_{xx}$ and the Nernst
signal $S_{xy}$ as functions of the Fermi energy in (a)-(b) biased
ABA-stacked trilayer graphene, and (c)-(d) biased ABC-stacked
trilayer graphene at different temperatures. The parameters in these
two systems are chosen to be the same as in Fig. 2 and Fig. 3,
respectively.} \label{fig.4}
\end{figure*}

We further calculate the thermopower $S_{xx}$ and the Nernst signal
$S_{xy}$. In Fig.\ref{fig.4}(a)-(b), we show the calculated $S_{xx}$
and $S_{xy}$ in biased ABA-stacked trilayer graphene. As we can see,
$S_{xy}$ ($S_{xx}$) has a peak (peaks) at the central LL (the other
LLs), and changes sign near the other LLs (the central LL). At zero
energy, both $\rho_{xy}$ and $\alpha_{xx}$ vanish, leading to a
vanishing $S_{xx}$. Around zero energy, because $\rho_{xx}
\alpha_{xx}$ and $\rho_{xy}\alpha_{xy}$ have opposite signs,
depending on their relative magnitudes, $S_{xx}$ could either
increases or decreases when $E_F$ is increased passing the Dirac
point. In biased ABA-stacked case, $S_{xx}$ is dominated by
$\rho_{xy}\alpha_{xy}$, consequently, $S_{xx}$ increases to positive
value as $E_F$ passing zero. At low temperatures, the peak value of
$S_{xx}$ near zero energy is $\pm 0.81 k_B/e$ ($\pm 69.8$ $\mu$V/K)
at $k_BT=0.4W_L$, which is in agreement with the measured value
($\pm 70$ $\mu$V/K).~\cite{LeiHao2010}
%%Here, the unit conversions is $k_B/e$ = 86.17
%%$\mu$V/K.
With the increase of temperature, the peak height increases to $\pm
3.75k_B/e$ ($\pm 323.14$ $\mu$V/K) at $k_BT=1.5W_L$. On the other
hand, $S_{xy}$ has a strong peak structure around zero energy, which
is dominated by $\rho_{xx}\alpha_{xy}$. The peak height is $7.82
k_B/e$ (673.85 $\mu$V/K) at $k_BT=1.5W_L$. The large magnitude of
$S_{xy}$ and $S_{xx}$ near zero energy can be attributed to the
semi-metal type dispersion of biased ABA-stacked trilayer graphene,
and the fact that the system is in the vicinity of a quantum Hall
liquid to insulator transition, where the imbalance between the
particle and hole types of carriers should be significant. The
thermoelectric effects are very sensitive to such an imbalance in
Dirac materials in comparison with conventional metals.

In Fig.\ref{fig.4}(c)-(d), we show the calculated $S_{xx}$ and
$S_{xy}$ in biased ABC-stacked case. As we can see, $S_{xy}$
($S_{xx}$) has a peak (peaks) around zero energy (the other LLs),
and changes sign near the other LLs (zero energy). These results are
qualitatively similar to those found in biased ABA-stacked case. In
our calculation, we find that $S_{xx}$ is always dominated by
$\rho_{xx}\alpha_{xx}$, consequently, $S_{xx}$ decreases to negative
values as $E_F$ passing zero. This is different from biased
ABA-stacked case. At low temperatures, the peak value of $S_{xx}$
near zero energy is $\pm 0.86k_B/e$ ($\pm 74.11$ $\mu$V/K) at
$k_BT=0.01\Delta_g$. With the increase of temperature, the peak
height increases to $\pm 2.23k_B/e$ ($\pm 192.16$ $\mu$V/K) at
$k_BT=0.1\Delta_g$. On the other hand, $S_{xy}$ has a peak structure
around zero energy, which is dominated by $\rho_{xy}\alpha_{xx}$.
The peak height is $4.69k_B/e$ (404.14 $\mu$V/K) at
$k_BT=0.1\Delta_g$.

\begin{figure*}
\par
\includegraphics[width=0.8\textwidth]{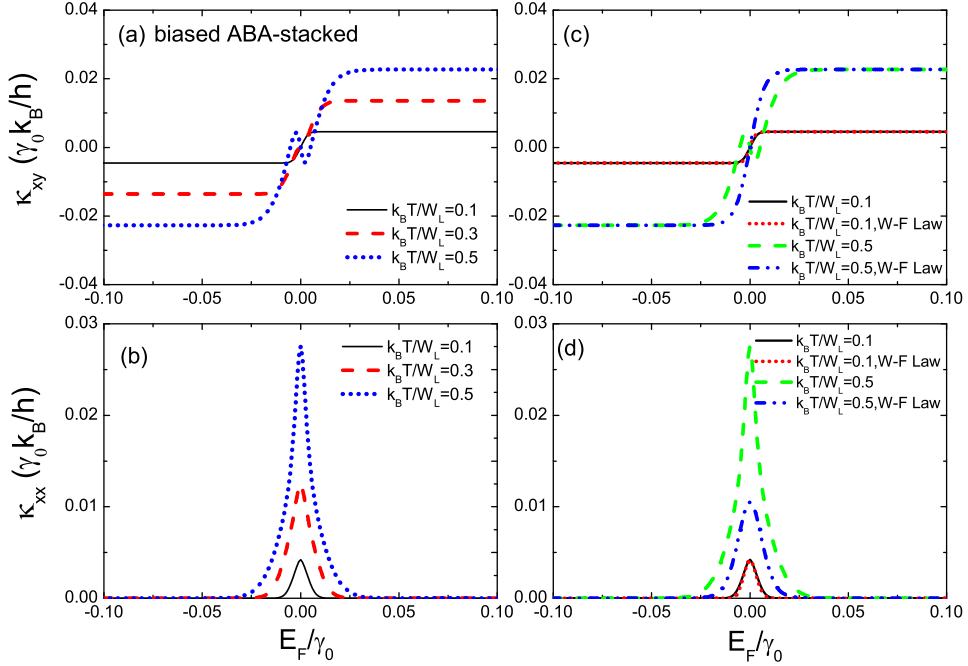}
\caption{(color online). (a)-(b) Thermal conductivities
$\kappa_{xy}(E_F,T)$ and $\kappa_{xx}(E_F,T)$ as functions of the
Fermi energy in biased ABA-stacked trilayer graphene at different
temperatures. (c)-(d) Comparison between the thermal conductivity as
functions of the Fermi energy from numerical calculations and from
the Wiedemann-Franz Law at two characteristic temperatures. The
parameters used here are the same as in Fig.2.} \label{fig.5}
\end{figure*}

\begin{figure*}
\par
\includegraphics[width=0.8\textwidth]{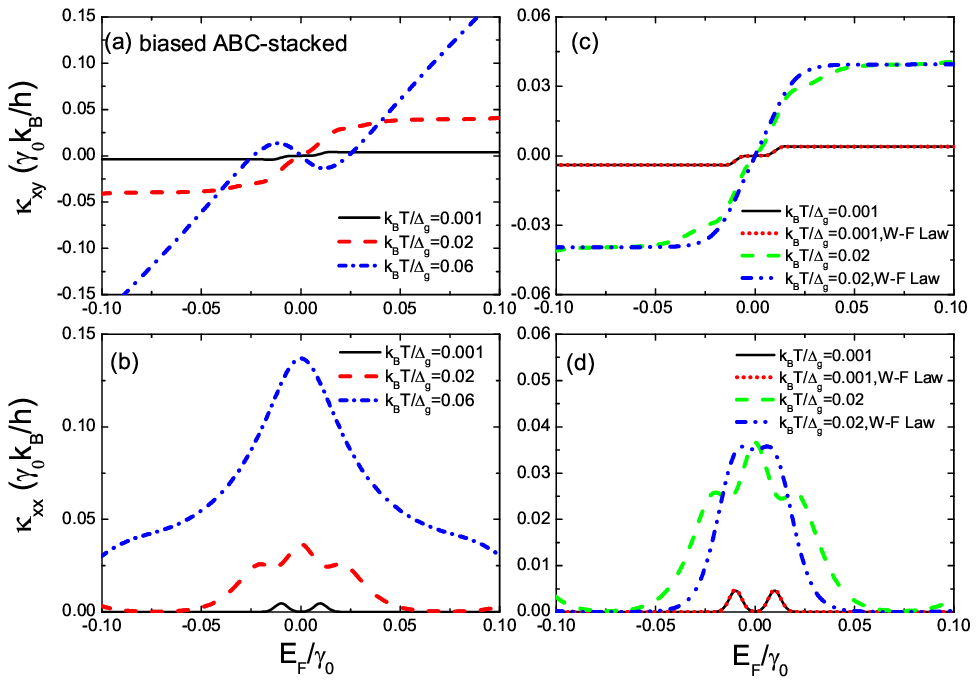}
\caption{ (color online). (a)-(b) Thermal conductivities
$\kappa_{xy}(E_F,T)$ and $\kappa_{xx}(E_F,T)$ as functions of the
Fermi energy in biased ABC-stacked trilayer graphene at different
temperatures, (c)-(d) Comparison between the thermal conductivity as
functions of the Fermi energy obtained from numerical calculations
and from the Wiedemann-Franz Law at two characteristic temperatures.
The parameters chosen here are the same as in Fig.3. } \label{fig.6}
\end{figure*}

\section{Thermal conductivity in biased ABA- and ABC-stacked trilayer graphene systems}
\label{eq:thermal}

We now focus on  thermal conductivities. In Fig.\ {\ref{fig.5}, we
show results of the transverse thermal conductivity $\kappa_{xy}$
and the longitudinal thermal conductivity $\kappa_{xx}$ for biased
ABA-stacked trilayer graphene at different temperatures. As seen
from Fig.\ref{fig.5}(a) and (b), $\kappa_{xy}$ exhibits two flat
plateaus away from the central LL. The values of the plateaus in
$\kappa_{xy}$ are $\pm 0.0045\gamma_0k_B/h$ ($\pm 0.048$
nW/(K$\cdot$m)) at $k_BT=0.1W_L$.
%%Here, the unit conversions is
%%$\gamma_0k_B/h$ = 10.62 nW/(K$\cdot$m).
With the increase of temperature, the values of the plateaus
increase to $\pm 0.023\gamma_0k_B/h$ ($\pm 0.24$ nW/(K$\cdot$m)) at
$k_BT=0.5W_L$. At low temperatures, the transition between these two
plateaus is smooth and monotonic, while at higher temperatures,
$\kappa_{xy}$ exhibits an oscillatory feature at $k_BT=0.5W_L$
between two plateaus. On the other hand, $\kappa_{xx}$ displays a
peak near the center LL, and its peak value increases quickly with
$T$. The peak height is $0.028\gamma_0k_B/h$ (0.3 nW/(K$\cdot$m)) at
$k_BT=0.5W_L$. To test the validity of the Wiedemann-Franz Law, we
compare the above results with those calculated from Eq.(\ref{eq:W-F
law}), as shown in Fig.\ref{fig.5}(c) and (d). The Wiedemann-Franz
Law predicts that the ratio of the thermal conductivity $\kappa$ to
the electrical conductivity $\sigma$ of a metal is proportional to
the temperature. This is in agreement with our low-temperature
results, but apparent deviation is seen at higher temperatures.

In Fig.\ {\ref{fig.6}, we show the calculated  thermal
conductivities $\kappa_{xx}$ and $\kappa_{xy}$ for biased
ABC-stacked case. As seen from Fig.\ref{fig.6}(a) and (b), around
zero energy, a flat region with $\kappa_{xy}=0$ is found at low
temperatures, which is accompanied by a valley in $\kappa_{xx}$.
These features are clearly in contrast to those of ABA-stacked case
due to the presence of an energy gap between the valence and
conduction bands. When temperature increases to $k_BT=0.06\Delta_g$,
the plateau with $\kappa_{xy}=0$ disappears, while $\kappa_{xx}$
displays a large peak. The peak height $\kappa_{xx}$ is
$0.14\gamma_0k_B/h$ (1.49 nW/(K$\cdot$m)) at $k_BT=0.06\Delta_g$. In
Fig.\ref{fig.6}(c) and (d), we also compare the above results with
those calculated from the Wiedemann-Franz Law using Eq.(\ref{eq:W-F
law}). We find that the Wiedemann-Franz Law remain valid at low
temperatures.

\section{Summary}
\label{sec:sum}

In summary, we have numerically investigated the thermoelectric and
thermal transport properties of biased trilayer graphene with
different stacking orders in the presence of both disorder and a
strong magnetic field. In biased ABA-stacked case, the
thermoelectric coefficients exhibit unique characteristics due to
the LL crossing of electron and hole bands that are strongly
suggestive of a semi-metallic band overlap. We find that the
thermoelectric conductivities display different asymptotic behavior
depending on the ratio between the temperature and the width of the
disorder-broadened LLs, similar to those found in monolayer
graphene. In the high temperature regime, the transverse
thermoelectric conductivity $\alpha_{xy}$ saturates to a universal
value $2.77 k_B e/h$ at the center of each LL, and displays a linear
temperature dependence at low temperatures. The calculated Nernst
signal $S_{xy}$ shows a strong peak at the central LL with heights
of the order of $k_B/e$, and changes sign at the other LLs, while
the thermopower $S_{xx}$ has an opposite behavior. The calculated
transverse thermal conductivity $\kappa_{xy}$ exhibits two plateaus
away from the band center. The transition between these two plateaus
is continuous, which is accompanied by a pronounced peak in
longitudinal thermal conductivity $\kappa_{xx}$. The validity of the
Wiedemann-Franz law relating the thermal conductivity $\kappa$ and
the electrical conductivity $\sigma$ is verified to be valid only at
very low temperatures.

In biased ABC-stacked case, the thermoelectric coefficients display
quite distinct behaviors from those of ABA-stacked case. Around the
Dirac point, the transverse thermoelectric conductivity
$\alpha_{xy}$ exhibits a pronounced valley with $\alpha_{xy}=0$ at
low temperatures, in striking contrast to ABA-stacked case with a
peak. The validity of the semiclassical Mott relation between the
thermoelectric and electrical transport coefficients is verified to
be satisfied only at very low temperatures. Furthermore, the
transverse thermal conductivity $\kappa_{xy}$ has a pronounced
plateau with $\kappa_{xy}=0$, which is accompanied by a valley in
$\kappa_{xx}$. These are consistent with the opening of sizable gap
between the valence and conductance bands in biased ABC-stacked
case.

We mention that in our numerical calculations, the flux $2\pi/M$ in
each hexagon gives a magnetic field of the strength $B \sim
1.3\times 10^5/M$ Tesla~\cite{Bernevig06}. Thus the magnetic field
$B$ we used is about $2700$ Tesla. This magnetic field is much
stronger than the ones which can be achieved in the experimental
situation, as limited by current computational capability. In our
calculation, the system size is taken to be $N=96\times24\times3$,
and $M$ is taken to be $L_x$ or $L_y$ in consistence with periodic
boundary conditions, which limits us to extremely strong magnetic
fields. However, the obtained thermoelectric transport coefficients
exhibit universal behaviors, as long as $M$ is not too small
(greater than 10).

\acknowledgments This work is supported by Scientific Research
Foundation of Nanjing University of Information and Technology of
China under Grant No. 20100401, the NSFC Grant No. 11104146 (RM),
the NSFC Grant No. 11074110, the National Basic Research Program of
China under Grant No 2009CB929504 (LS). We also thank the US NSF
Grants DMR-0906816 and DMR-1205734 (DNS), Princeton MRSEC Grant
DMR-0819860 for travel support, and the NSF instrument grant
DMR-0958596 (DNS).


\begin{references}
\bibitem{footnote1} Different literatures may have a sign difference due to different conventions.
\end{references}
\end{document}